\documentclass{article}
\usepackage{moreverb}
\usepackage{tikz}
\usepackage[numbers]{natbib}
\usepackage{booktabs}
\usepackage{amsmath}
\usetikzlibrary{arrows.meta}
\usepackage{graphicx} %% for Jpegs\usetikzlibrary{arrows.meta}s
\usetikzlibrary{shapes,positioning}
\begin{document}

\title{Futility analyses for the MCP-Mod methodology based on longitudinal models}

\date{}
\author{Björn Bornkamp\thanks{Advanced Methodology and Data Science, Novartis Pharma AG, Basel, Switzerland}
\and Jie Zhou\thanks{Neuroscience Biostatistics, Novartis Pharmaceuticals Corporation, East Hanover, New Jersey, USA}
\and Dong Xi\thanks{Department of Biostatistics, Gilead Sciences, Foster City, California, USA}
\and Weihua Cao\footnotemark[2]
}

\maketitle

\abstract{This article discusses futility analyses for the MCP-Mod methodology. Formulas are derived for calculating predictive and conditional power for MCP-Mod, which also cover the case when longitudinal models are used allowing to utilize incomplete data from patients at interim. A simulation study is conducted to evaluate the repeated sampling properties of the proposed decision rules and to assess the benefit of using a longitudinal versus a completer only model for decision making at interim. The results suggest that the proposed methods perform adequately and a longitudinal analysis outperforms a completer only analysis, particularly when the recruitment speed is higher and the correlation over time is larger. The proposed methodology is illustrated using real data from a dose-finding study for severe uncontrolled asthma. 
}

\section{Introduction}
\label{sec:intro}

A futility analysis is a planned interim data assessment to determine whether it is unlikely for the trial to achieve its primary objective at trial end, which can lead to an early termination of the trial. It is a risk mitigation strategy to protect participants from unnecessary exposure to ineffective treatments and to conserve resources. Futility interim analyses are conducted routinely in Phase 3 clinical trials with comprehensive reviews available\cite{jennison2000group, lachin2005review, snapinn2006assessment, proschan2006statistical, gallo2014alternative, ellenberg2019data}. However, the failure rate of Phase 2 trials is even higher. For data from 2003 to 2011 only a rate of around 32\% of the drugs transitioned from Phase 2 to Phase 3 of clinical development\cite{hay2014clinical}. For the time frame of 2013-2015 the most commonly reported reason for failure in Phase 2 was lack of efficacy in around 48\% of the cases\cite{harrison2016phase}. These data suggest that futility analyses should be an important consideration of clinical trial design also in Phase 2 trials. 

A commonly used design and analysis strategy for randomized Phase 2 dose-finding studies is the Multiple Comparisons and Modelling (MCP-Mod) approach, which allows for testing of the existence of a non-flat dose-response relationship versus placebo and modelling the dose-response function, while acknowledging model uncertainty\cite{bret:pinh:bran:2005, pinh:born:bret:2006, pinh:born:glim:2014}. It consists of two steps: in the first step a multiple contrast test is used to test whether there is evidence against the null hypothesis that all treatment groups have the same mean, while controlling the familywise error rate. In these tests the contrasts are derived so that they are efficient for detecting dose-response shapes corresponding to pre-specified candidate models. The second step consists of estimating the dose-response relationship, often by using model averaging over the assumed candidate dose-response models. The first step is typically the primary objective of the trial and can be seen as a "gatekeeper" or minimum requirement in terms of efficacy before estimating the dose-response relationship and potential target doses of interest. 

Many statistical metrics have been proposed to describe the possible interim decision to stop a clinical trial due to futility. Concepts such as the conditional power\cite{lachin2005review} and the predictive power\cite{spiegelhalter1986monitoring} are popular choices for futility boundaries based on statistical curtailment\cite{gordon1982stochastically}. An alternative approach applies a group sequential design in a similar way for efficacy interim analyses to derive futility boundaries using $z$-value or $p$-value\cite{pampallona1994group}. Inter-relationships among these futility boundaries have been established and it has been suggested to design futility interim analyses based on operating characteristics, e.g., the probability of stopping when there is no treatment effect and the power loss due to incorrectly stopping when there is a positive treatment effect\cite{emerson2007frequentist, gallo2014alternative}. Timing of futility interim analyses have also been investigated\cite{pallay2000timing, xi2017optimal}.

Utilizing incomplete measurements from patients at interim to increase the efficiency of decision making at interim analyses for sample size re-estimation or futility analysis has been proposed by a few authors recently see for example\cite{hampson2013group, van2020improving}. The authors focused primarily on the situation of two-arm trials. In the context of MCP-Mod and multiple contrast tests more generally formulas were described recently for conditional and predictive power calculations that could equally be used in the setting of futility calculations\cite{liu2023sample}. However these authors do not consider the situation of using incomplete data from patients at interim.

The purpose of this paper is to derive interim decision metrics for futility analyses when using MCP-Mod, while utilizing longitudinal data. We focus on futility assessments, but the methods could equally be used for unblinded sample size re-estimation. In Section \ref{sec:intro_example} we outline a recent example of a dose-finding study to illustrate the proposed methodology. Section \ref{sec:methodology} describes the underlying methodology. A simulation study is conducted to evaluate the repeated sampling properties of the approach and to characterize when most gain can be expected from a longitudinal approach that uses all data versus an  alternative approach that only uses completer data at interim. Then in Section \ref{sec:example} we return to the example and calculate the proposed rules for the study presented in Section \ref{sec:intro_example}. Section \ref{sec:concl} provides a discussion and conclusion.

\section{Dose-finding study example}
\label{sec:intro_example}

For illustration purposes we consider a recent dose-finding study for severe uncontrolled asthma that investigated five doses (0.5, 1, 2, 4 and 8mg) of a new investigational treatment versus placebo (clinicaltrials.gov identifier NCT04410523). A randomization ratio of 2:1:1:1:2:2 was used for placebo and the investigational doses and it was planned to recruit around 630 patients in total. Randomization was stratified by the baseline eosinophil (EOS) count ($\geq$ 300 or < 300 cells/µl). The clinical outcome of interest is forced expiratory volume in one second (FEV1), which measures the volume of air which can be forcibly exhaled from the lungs in the first second of exhalation in a spirometry test. This outcome was measured from all patients at the baseline visit and at all post-baseline visits at week 2, 4, 8 and 12. The primary endpoint for this study was the average change from baseline in pre-dose FEV1 at Week 8 and Week 12. The primary anticipated intercurrent event, study treatment discontinuation, was planned to be handled with a hypothetical strategy\cite{ICH2019}. The study was stopped pre-maturely after recruiting around half of the patients, due to a business decision independent of the observed data in the trial.

\begin{figure}
    \centering
    \includegraphics[width=0.6\linewidth]{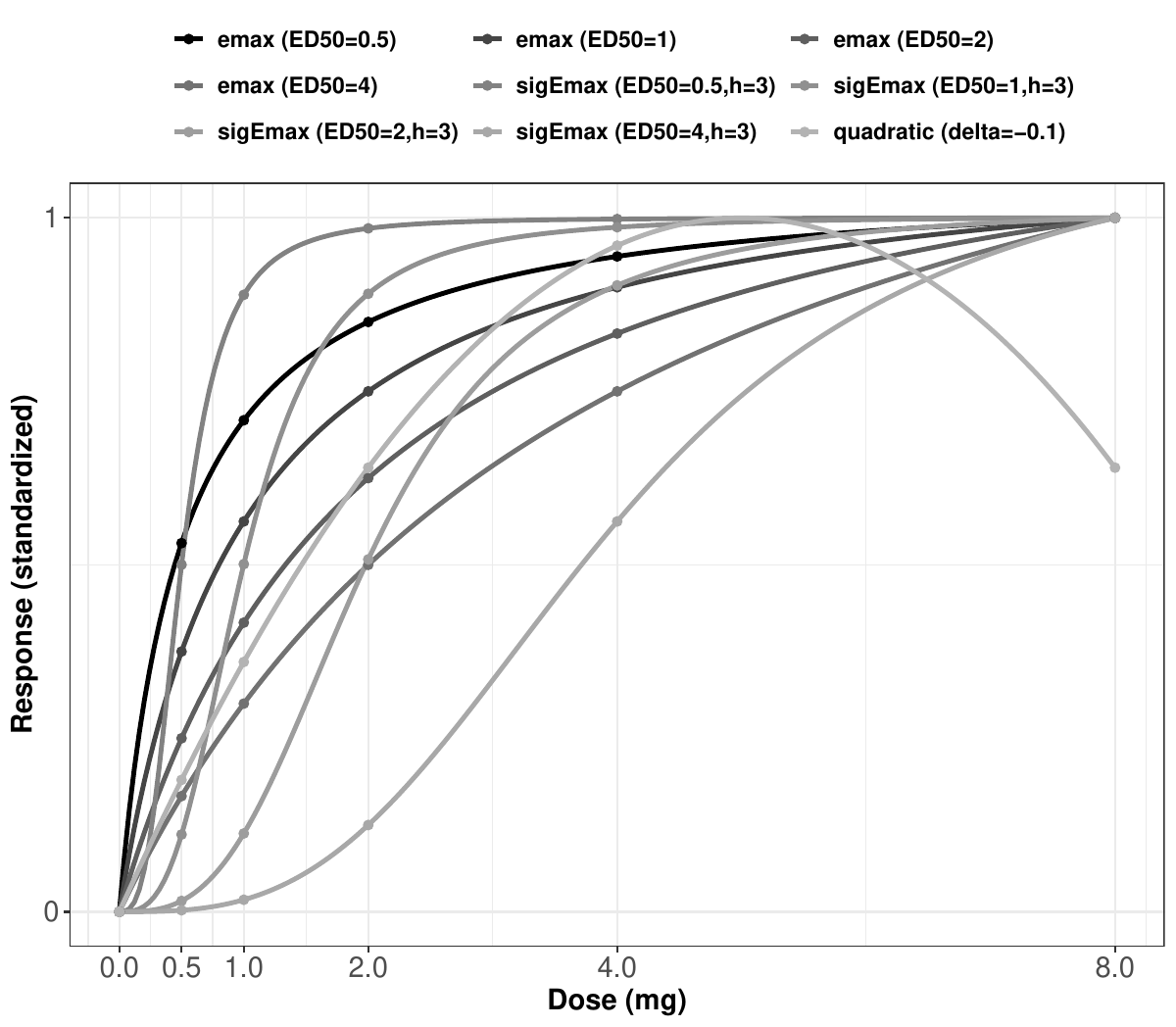}
    \caption{Candidate dose-response shapes pre-specified for dose-finding study example.}
    \label{fig:cand_mod}
\end{figure}

Generalized MCP-Mod\cite{pinh:born:glim:2014} was planned as the primary analysis in this trial: A mixed model repeated measures (MMRM) model\cite{siddiqui2009mmrm} was pre-specified to be fitted to FEV1 change from baseline  including all post-baseline visits at week 2, 4, 8 and 12. The model includes the following fixed effects: treatment, randomization strata (EOS $\geq$ 300 or < 300 cells/µl), region, treatment-by-visit interaction, baseline FEV1-by-visit interaction, FEV1 prior to inhalation, and FEV1 within 30 min post inhalation of salbutamol/albuterol (components of SABA reversibility). To allow adjustment for correlations between time points within patients, an unstructured variance-covariance structure was planned to be used. Then based on this model adjusted population means for the endpoint of interest were planned to be extracted for each treatment group. Generalized MCP-Mod was then planned to be performed on these adjusted means for change from baseline averaged between Week 8 and 12. The longitudinal MMRM model was used, because it targets the hypothetical strategy and utilizes patients in the analysis, who have incomplete data (e.g. missing Week 8 or 12, but available earlier measurements), under the principled approach of a missing at random assumption. At the design stage of the trial, four Emax, four sigmoid Emax and one quadratic dose-response shape were pre-specified as the candidate model set (see Figure \ref{fig:cand_mod}). The Emax and sigmoid Emax candidate models correspond to monotonic dose-response shapes with different steepness and the quadratic model assumes the response will first increase and then decrease within the observed dose-range.

The results of the study are available on clinicaltrials.gov and not further discussed here. This example will serve as a motivation for the considered simulation scenarios later in Section \ref{sec:sims} and we will return to this example in Section \ref{sec:example} to  retrospectively perform futility analyses based on the methodologies proposed in the following section. 

\section{Methodology}
\label{sec:methodology}

In this section we will derive formulas on how to calculate the conditional power or the predictive power at an interim analysis, when using the MCP-Mod methodology. For that purpose we will first consider the setting of a single visit (with no longitudinal data) and a normally distributed homoscedastic endpoint\cite{bret:pinh:bran:2005}. Then we will extend this to the setting of generalized MCP-Mod\cite{pinh:born:glim:2014}. Generalized MCP-Mod is a two-stage procedure: In the  first stage a model is used to derive treatment group specific adjusted estimates and their associated covariance matrix. Then in the second stage MCP-Mod is directly conducted on these summary estimates. While generalized MCP-Mod applies to many different situations (for example to approximate inference for non-normal endpoints), we are specifically interested here in the setting of using longitudinal models. They have the advantage of utilizing all data from patients that may have missing outcome data at a particular time-point of interest, which is of particular interest for decision making at interim analyses. 

\subsection{Homoscedastic normally distributed outcome}

Let $d_1,...,d_k$ be the different doses with $d_1$ being the placebo, and $n_1,...,n_k$ the sample sizes at the different doses. We assume the outcome $Y_{ij}\sim N(\mu_i,\sigma^2)$, $i=1,...,k$ and $j=1,...,n_i$, with $\sigma$ known and $\boldsymbol{\mu}=(\mu_1,...,\mu_k)'$ the true underlying population group means.

Let $\textbf{c}_1,...,\textbf{c}_M$ be the contrast vectors underlying the $M$ candidate models. and $T_1,...,T_M$ the corresponding contrast test statistics defined by $$T_m=\frac{\textbf{c}_m'\bar{\boldsymbol{Y}}}{\sigma\sqrt{\sum_{i=1}^k\frac{c_{m,i}^2}{n_i}}}.$$ A critical value $c_{1-\alpha}$ is determined so that $P(\max_{m \in \{1,...,M\}}T_m > c_{1-\alpha})=1-P(T_1<c_{1-\alpha},...,T_M<c_{1-\alpha})=\alpha$ under the null-hypothesis $\mu_1=...=\mu_k$ based on numerical integration for the multivariate normal distribution\cite{bret:pinh:bran:2005, pinh:born:bret:2006, pinh:born:glim:2014}.

The vector of test statistics $\textbf{T}=(T_1,...,T_M)'$ is given by $\textbf{T}=\textbf{P}\textbf{C}\bar{\boldsymbol{Y}}$, where $\textbf{C}$ is the $M\times k$ matrix with the contrast vectors $\textbf{c}'_m$ in the $m$-th row, and $\textbf{P}$ is the $M\times M$ dimensional diagonal matrix with the $m$-th diagonal element being equal to $\left(\sigma\sqrt{\sum_{i=1}^k\frac{c_{m,i}^2}{n_i}}\right)^{-1}$.

In what follows we will denote by $\bar{Y}_{i,[0,t]}$ the group mean for group $i$ at the information time $t$, which is also called the information fraction and equal to the proportion of patients with observed outcome $Y$. Time $t=1$ would correspond to the group mean at the final analysis. Similarly, let $\bar{Y}_{i,(t,1]}$ denote the group means of patients between time $t$ and end of study. It can be shown that the variance of $\bar{Y}_{i,[0,t]}$ and $\bar{Y}_{i,(t,1]}$ are $\sigma^2/(tn_{i})$ and $\sigma^2/((1-t)n_{i})$, respectively.

Assume an interim analysis at time $t$. The variance at study end is $\sigma^2/n_i$ for dose group $i$. The overall vector of group means at study end is given by

\begin{equation}
    \label{eqn:overallmean}
    \bar{Y_i}=(tn_i\bar{Y}_{i,[0,t]}+(1-t)n_i\bar{Y}_{i,(t,1]})/n_i=t\bar{Y}_{i,[0,t]}+(1-t)\bar{Y}_{i,(t,1]}^{(b)}
\end{equation}
with the variance of $\sigma^2 t/n_{i}+\sigma^2 (1-t)/n_{i}=\sigma^2/n_{i}$. In vector form the group mean is $t\bar{\boldsymbol{Y}}_{[0,t]}+(1-t)\bar{\boldsymbol{Y}}_{(t,1]}$ so that the vector of test statistics at study end can be written as

\begin{eqnarray}
  \label{eqn:teststat}
  \textbf{T}=\textbf{P}\textbf{C}(t\bar{\boldsymbol{Y}}_{[0,t]}+(1-t)\bar{\boldsymbol{Y}}_{(t,1]}).
\end{eqnarray}

At the interim analysis the vector $\bar{\boldsymbol{Y}}_{[0,t]}$ has already been observed, but $\bar{\boldsymbol{Y}}_{(t,1]}$ is unknown. It is of interest to assess whether $\max_{m \in \{1,...,M\}}T_m > c_{1-\alpha}$ at study end. The sampling distribution of $\bar{\boldsymbol{Y}}_{(t,1]}$ for the remaining patients is a $MVN(\boldsymbol{\mu},\sigma^2(1-t)^{-1}\textbf{D}))$ distribution, where $\textbf{D}$ is a $k\times k$ diagonal matrix with i-th diagonal entry equal to $1/n_i,\, i=\{1,...,k\}$. 

We consider three different ways to deal with the uncertainty in $\boldsymbol{\mu}$: Predictive power and two variants of the conditional power. In a Bayesian setting the posterior distribution for $\boldsymbol{\mu}$, using an improper constant prior, is a normal-distribution $MVN(\bar{\boldsymbol{Y}}_{[0,t]},\sigma^2t^{-1}\textbf{D})$. This implies that the Bayesian predictive distribution for $\bar{\boldsymbol{Y}}_{(t,1]}$ is equal to $N(\bar{\boldsymbol{Y}}_{[0,t]}, \sigma^2t^{-1}\textbf{D} + \sigma^2(1-t)^{-1}\textbf{D})$. This distribution can be used to calculate the predictive power. 

Another approach is to assume a constant ``guess'' $\tilde{\boldsymbol{\mu}}$ for $\boldsymbol{\mu}$. Then the distribution is $N(\tilde{\boldsymbol{\mu}},\sigma^2(1-t)^{-1}\textbf{D})$. This can be used for conditional power calculation. We will consider two variants of conditional power, where $\tilde{\boldsymbol{\mu}}$ is chosen either to be equal to the design assumption or based on the estimate based on the first stage.

Based on this the probability $P(\max_{m \in \{1,...,M\}}T_m > c_{1-\alpha})$ at study end can be calculated based on the equi-coordinate distribution function of the multivariate normal distribution using numerical integration. Based on \eqref{eqn:teststat} the multivariate normal distribution for the Bayesian predictive distribution has mean $\textbf{P}\textbf{C}(t\bar{\boldsymbol{Y}}_{[0,t]}+(1-t)\bar{\boldsymbol{Y}}_{[0,t]}))=\textbf{P}\textbf{C}\bar{\boldsymbol{Y}}_{[0,t]}$ and variance $\textbf{P}\textbf{C}(\sigma^2(1-t)^2(t^{-1}+ (1-t)^{-1})\textbf{D})\textbf{C}'\textbf{P}'$, while for conditional power calculation we obtain mean $\textbf{P}\textbf{C}(t\bar{\boldsymbol{Y}}_{[0,t]}+(1-t)\tilde{\boldsymbol{\mu}})$ and variance $\textbf{P}\textbf{C}(\sigma^2(1-t)\textbf{D})\textbf{C}'\textbf{P}'$.

\subsection{Generalized MCP-Mod}

For the setting of generalized MCP-Mod, we assume that the adjusted dose group estimates at any specific time-point of interest are multivariate normally distributed with a given covariance matrix. The adjusted dose group estimates $\hat{\boldsymbol{\mu}}$ at study end are assumed to be multivariate normally distributed with mean $\boldsymbol{\mu}$ and covariance matrix $\textbf{S}$.

The contrast test statistics $T_1,...,T_M$ corresponding to the $M$ candidate shapes are now defined by $$T_m=\frac{\textbf{c}_m'\hat{\boldsymbol{\mu}}}{\sqrt{\textbf{c}_m'\textbf{S}\textbf{c}_m}}.$$
The vector of test statistics $\textbf{T}=(T_1,...,T_M)'$ is given by $\textbf{T}=\textbf{P}\textbf{C}\hat{\boldsymbol{\mu}}$, where $\textbf{C}$ is the $M\times k$ matrix with the contrast vectors $\textbf{c}'_m$ in the m-th row, and $\textbf{P}$ is the $M\times M$ dimensional diagonal matrix with the m-th diagonal entry being equal to $(\textbf{c}_m'\textbf{S}\textbf{c}_m)^{-1/2}$.

In what follows we will denote by $\hat{\boldsymbol{\mu}}_{[0,t]}$ the interim estimates for the visit of interest at time $t \in [0,1]$ (for example based on a longitudinal model) with corresponding covariance matrix $\textbf{S}_{[0,t]}$.  Similarly let $\hat{\boldsymbol{\mu}}_{(t,1]}$ and $\textbf{S}_{(t,1]}$ be the corresponding quantities based on the time between $t$ and the end of trial. Here we assume that $\textbf{S}_{[0,t]}$ and $\textbf{S}_{(t,1]}$ are known and that $\hat{\boldsymbol{\mu}}_{[0,t]}\sim MVN(\boldsymbol{\mu}, \textbf{S}_{[0,t]})$ and $\hat{\boldsymbol{\mu}}_{(t,1]}\sim MVN(\boldsymbol{\mu}, \textbf{S}_{(t,1]})$. Finally $\textbf{S}$ will be denoted as $\textbf{S}_{[0,1]}$ to make clear that it is the covariance matrix based on all data. In addition note that $\hat{\boldsymbol{\mu}}_{[0,t]}$ and $\hat{\boldsymbol{\mu}}_{(t,1]}$ are independent, which also holds in the longitudinal setting, when there are partially observed patients, i.e. data from the same patient are ``part'' of both stages, see Section 3.4 in \cite{jennison2000group}.

In the previous section the final estimate is written as a weighted average of the stage-specific estimates. In the generalized situation considered here, we have one observation per stage from a multivariate normal distribution, so two observations in total. When multiplying the multivariate normal densities underlying these two observations, one again obtains a multivariate normal density with parameters given by
$\textbf{S}_{[0,1]}= (\textbf{S}_{[0,t]}^{-1}+\textbf{S}_{(t,1]}^{-1})^{-1}$ and $\hat{\boldsymbol{\mu}}=\textbf{S}_{[0,1]}(\textbf{S}_{[0,t]}^{-1}\hat{\boldsymbol{\mu}}_{[0,t]}+\textbf{S}_{(t,1]}^{-1}\hat{\boldsymbol{\mu}}_{(t,1]})$ (see \cite[Chapter 14.3]{ohag:fors:2004}). For the special case of homoscedastic normally distributed outcomes, this formula reduces to the overall vector of group means at the study end as in \eqref{eqn:overallmean}.

Assume an interim analysis at time $t$. The overall estimate at study end can thus be written as $\hat{\boldsymbol{\mu}}=\textbf{S}_{[0,1]}(\textbf{S}_{[0,t]}^{-1}\hat{\boldsymbol{\mu}}_{[0,t]}+\textbf{S}_{(t,1]}^{-1}\hat{\boldsymbol{\mu}}_{(t,1]})$. This implies that the vector of test statistics at study end is given by

\begin{eqnarray}
  \label{eqn:teststat_gen}
  \textbf{T}=\textbf{P}\textbf{C}(\textbf{S}_{[0,1]}(\textbf{S}_{[0,t]}^{-1}\hat{\boldsymbol{\mu}}_{[0,t]}+\textbf{S}_{(t,1]}^{-1}\hat{\boldsymbol{\mu}}_{(t,1]})).
\end{eqnarray}

At the interim analysis the vector $\hat{\boldsymbol{\mu}}_{[0,t]}$ has already been observed, but $\hat{\boldsymbol{\mu}}_{(t,1]}$ is unknown. It is of interest to assess whether $\max_{m \in \{1,...,M\}}T_m > c_{1-\alpha}$ at study end. The sampling distribution for the remaining observations is a $MVN(\boldsymbol{\mu}, \textbf{S}_{(t,1]})$ distribution. 

We again consider predictive and conditional power. In a Bayesian setting (for predictive power calculation) the posterior distribution for $\boldsymbol{\mu}$, using an improper constant prior, is a normal-distribution $MVN(\hat{\boldsymbol{\mu}}_{[0,t]}, \textbf{S}_{[0,t]})$. Together with the sampling distribution above, this implies that the Bayesian predictive distribution for $\hat{\boldsymbol{\mu}}_{(t,1]}$ is equal to $MVN(\hat{\boldsymbol{\mu}}_{[0,t]},\textbf{S}_{[0,t]}+\textbf{S}_{(t,1]})$. 

For conditional power calculation if a constant ``guess'' $\tilde{\boldsymbol{\mu}}$ is used for $\boldsymbol{\mu}$, the distribution would then be $MVN(\tilde{\boldsymbol{\mu}}, \textbf{S}_{(t,1]})$, where commonly $\tilde{\boldsymbol{\mu}}$ is chosen either to be equal to the design assumption or based on the estimate based on the first stage.

As in the previous section $P(\max_{m \in \{1,...,M\}}T_m > c_{1-\alpha})=1-P(T_1<c_{1-\alpha},...,T_M<c_{1-\alpha})$ at study end can be calculated based on the equi-coordinate distribution function of the multivariate normal distribution using numerical integration. Based on \eqref{eqn:teststat_gen} and applying rules for calculating the distribution of linear transformations of multivariate normal distributions the (conditional) multivariate normal distribution for the Bayesian predictive distribution of the vector of teststatistics has mean $\textbf{P}\textbf{C}\hat{\boldsymbol{\mu}}_{[0,t]}$ and variance $\textbf{P}\textbf{C}\textbf{S}_{[0,1]}\textbf{S}_{(t,1]}^{-1}(\textbf{S}_{[0,t]}+\textbf{S}_{(t,1]})\textbf{S}_{(t,1]}^{-1}\textbf{S}_{[0,1]}\textbf{C}'\textbf{P}'$, while for the conditional power calculation we obtain a mean $\textbf{P}\textbf{C}\textbf{S}_{[0,1]}(\textbf{S}_{[0,t]}^{-1}\hat{\boldsymbol{\mu}}_{[0,t]}+\textbf{S}_{(t,1]}^{-1}\tilde{\boldsymbol{\mu}})$ and variance $\textbf{P}\textbf{C}\textbf{S}_{[0,1]}\textbf{S}_{(t,1]}^{-1}\textbf{S}_{[0,1]}\textbf{C}'\textbf{P}'$.

In the previous section the information fraction, due to the homoscedastic normal setting is just the fraction of outcomes included in the analysis. For the generalized setting we propose to define the information fraction as the ratio of the reciprocal of determinant of the covariance matrix at the interim analysis and at the end of study to the power of $1/k$, i.e.
\begin{eqnarray}
\label{eqn:det}
t = \frac{\det(\mathbf{S}_{[0,1]})^{1/k}}{\det(\mathbf{S}_{[0,t]})^{1/k}},
\end{eqnarray}
where $\det(\cdot)$ is the determinant function of a matrix. Similar criteria are used under the term D-efficiency in the optimal design literature\cite{dette2010optimal}. If the covariance matrices are diagonal, the determinants above would reduce to the geometric mean of the variances of the adjusted mean estimates.

While the results in this section apply for generalized MCP-Mod (i.e. different selection of first stage models), our primary interest in this paper is when a longitudinal model is used at interim to estimate $\hat{\boldsymbol{\mu}}_{[0,t]}$. In this setting the incomplete information of patients in the trial, who have not yet completed the final endpoint visit is incorporated in $\hat{\boldsymbol{\mu}}_{[0,t]}$ and the covariance matrix $\textbf{S}_{[0,t]}$. The determination of the information fraction \eqref{eqn:det} at interim timing $t$ thus includes all patients recruited by time $t$. So despite the fact that some patients may have missing outcome data at the time-point of interest, their outcome data from other time-points contribute to the information fraction. 

Note that the results in this section only hold approximately in the practically relevant case, where the covariance matrices $\textbf{S}_{[0,t]}$, $\textbf{S}_{(t,1]}$ and $\textbf{S}_{[0,1]}$ are estimated. Estimation of these matrices in the longitudinal setting would proceed as follows: $\textbf{S}_{[0,t]}$ would be estimated by the longitudinal model itself. But also estimates of $\textbf{S}_{[0,1]}$ as well as $\textbf{S}_{(t,1]}$ are needed in the formulas for predictive and conditional power. Here one would estimate $\textbf{S}_{[0,1]}$ either purely based on the design assumptions regarding the endpoint variability, or by using the known structure for $\textbf{S}_{[0,1]}$ and estimating the residual variance based on the interim analysis data. An estimate for $\textbf{S}_{(t,1]}$ can then be determined by using the equation $\textbf{S}_{[0,1]}= (\textbf{S}_{[0,t]}^{-1}+\textbf{S}_{(t,1]}^{-1})^{-1}$, plugging in estimates for $\textbf{S}_{[0,t]}$ and $\textbf{S}_{[0,1]}$ and solving for $\textbf{S}_{(t,1]}$.

\section{Evaluation of benefit of longitudinal versus completer futility analyses}
\label{sec:benefit}

A futility analysis using all longitudinal data will contain more information than an analysis based on the patients that completed the primary endpoint visit at the interim analysis (which we will call "completer analysis" in what follows). In this section we will investigate the factors that impact the information gain for a longitudinal analysis. Basic considerations suggest that a longitudinal analysis will outperform a completer analysis, when (i) there are many patients at interim with incomplete data (which are not used by the completer analysis) and (ii) when earlier endpoints are more predictive of the final endpoint. In Appendix \ref{sec:illus} we provide an illustration for a simplified situation. There it becomes clear that a major factor for point (i) is the recruitment speed of the trial, relative to the time it takes to obtain the final measurement for one patient. If the recruitment is faster, relative to the time it takes to obtain the final measurement for one patient, there will be a higher proportion of incomplete patients at interim and a longitudinal approach is expected to outperform a completer only approach. For (ii) the correlation of outcomes over time within a patient plays an important role. Different scenarios are designed in the simulation in terms of these two respects to evaluate the impact on the information fraction and results.

\subsection{Simulations}
\label{sec:sims}

\subsubsection{Design of Simulation Study}

Following the ADEMP framework\cite{morris2019using}, we define aims, data-generating mechanism, methods and performance measures for the simulation study.\\ 

\noindent \textbf{Aims}\\
The aim of this simulation study is to evaluate the repeated sampling properties of the proposed decision rules based on the models presented in Section \ref{sec:methodology}. Another aim is to assess the benefit of using a longitudinal versus a completers only model for decision making at interim.\\

\noindent \textbf{Data generating mechanisms}\\
Data generation is motivated by the example in Section \ref{sec:intro_example} where participants are randomized into six arms (placebo, active doses 0.5mg, 1mg, 2mg, 4mg and 8mg) with ratio 2:1:1:1:2:2.
Recruitment time ($T^{rec}$) of the participants is assumed to follow a quadratic shape $T^{rec}_i=T^{rec}_{N}\sqrt{U_i}$ for $i=1, \cdots, N$, where $T^{rec}_{N}$ is the last patient first visit time (LPFV.t) and $U_i$ is randomly generated from standard uniform distribution. With fixed total sample size $N$, the time of last patient first visit determines the recruitment speed and plays a crucial role on the information fraction. Therefore, we consider two different scenarios $LPFV.t=50$ or $100$ weeks corresponding to cases with faster or slower recruitment.

Outcome data were simulated from a multivariate normal distribution for visits at baseline, week 2, 4, 8 and 12. Two scenarios for the maximum effect are assumed: the null case with flat dose-response relationship versus placebo (zero treatment effect) and the planned maximum effect of 0.12 L. When the true dose-response is not flat, the mean responses at each dose level are assumed to follow the Emax function $E_0+E_{max}(T)\frac{d}{d+ED_{50}}$ with placebo effect $E_0=0$ and $ED_{50}=1$, here $d$ denotes the dose and $T$ the time in weeks. The dose-response relationship is assumed to be monotone and the maximum response 0.12 is achieved in the highest dose arm (8mg) and at the last visit (week 12). The maximum effect at specific time $T$ is derived as $$E_{max}(T)=E_{max}(T_{max})\times \frac{1-\exp\{-0.5\cdot T\}}{1-\exp\{-0.5\cdot T_{max}\}}, $$
where $T_{max}=12$ is the maximum visit time and $E_{max}(T_{max})=0.135$ is derived to achieve a maximum effect of 0.12 L for 8mg at week 12. The mean response of dose $d$ at time $T$ is then calculated as $E_{max}(T)\times d/(d+1)$.

The real data underlying the example in Section \ref{sec:intro_example} can be well approximated by a compound symmetry (CS) structure with standard deviation $\sigma=0.56$ and a correlation $\rho=0.9$. We consider using the similar structure in simulation to generate the error terms for the response. Specifically, we use the CS structure with $\sigma=0.56$ and consider two different scenarios $\rho=0.9$ and $0.6$ to investigate the effect of outcome correlation on the result.
Note that data generation is based on the outcome per patient, while the analysis at study end and at interim will use the change from baseline per patient, while adjusting for baseline in the analysis model.
The sample sizes for the scenarios corresponding to different $LPLV.t$ and $\rho$ are chosen so that the power at the final analysis is around 80\% when the true effect size is 0.12.
Timing of the interim analysis is determined when there are 30\%, 50\% or 70\% participants with completed week 12 assessment. With the data available at the interim analysis, the proposed methods are applied to calculate the predictive or conditional power based on the pre-specified candidate models in Figure \ref{fig:cand_mod}.
For each scenario, the results are summarized based on 5000 replications.\\

\noindent \textbf{Methods}\\
The predictive power and conditional power with interim data are calculated based on the methods in Section \ref{sec:methodology}. Under the generalized approach, the $\hat{\boldsymbol{\mu}}_{[0,t]}$ and $\boldsymbol{S}_{[0,t]}$ are derived using either a) the longitudinal data approach or b) the completer only approach. The longitudinal data approach utilizes all available longitudinal data (including partial data from incompleters) at interim and fit mixed models for repeated measures (MMRM) with unstructured covariance matrix and obtain the estimated least squares (LS) means for each dose at week 12 and the corresponding covariance matrix. On the other hand, the completer approach only uses the available outcome data at week 12 to derive the LS means estimates. In the conditional power calculation, the mean of the second stage data $\tilde{\boldsymbol{\mu}}$ is chosen to be equal to the estimated placebo group mean plus the average of the designed treatment effects at each of the doses from the candidate models in Figure \ref{fig:cand_mod} (using the planned maximum effect of 0.12). Additional results for the conditional power when $\tilde{\boldsymbol{\mu}}$ is set to the estimated means from the first stage $\hat{\boldsymbol{\mu}}_{[0,t]}$ are included in the Appendix \ref{sec:addres}.

Calculation of the predictive and conditional power requires the estimate of the covariance matrix $\boldsymbol{S}_{[0,1]}$ at study end. It is known that for complete data $\boldsymbol{S}_{[0,1]}$ is a diagonal matrix with the $i^{th}$ element being $\hat\sigma^2/n_i$. So the standard deviation $\hat\sigma$ is estimated at interim in both models based on the interim data. The covariance matrix for the future patients $\boldsymbol{S}_{[t,1]}$ can be derived as $(\boldsymbol{S}_{[0,1]}^{-1}-\boldsymbol{S}_{[0,t]}^{-1})^{-1}$.

The contrast matrix $\boldsymbol{C}$ and the corresponding critical value based on the candidate models can be obtained using the ``DoseFinding'' R package\cite{DoseFinding}. Based on the approximated multivariate normal distribution, the predictive and conditional powers can be derived using the `pmvnorm' function in the ``mvtnorm'' R package\cite{mvtnorm}.\\

\noindent \textbf{Performance Measures}\\

The performances of the two approaches are compared in both the null setting where no treatment effect is present for any of the investigational doses and in the alternative setting where dose-response curve is not flat and the maximum effect of 0.12 is achieved at the highest dose. To show the gain of using all available longitudinal data compared with only using the completers' data in the interim analysis, we use the information fraction defined in \eqref{eqn:det}.

The cut-offs for decision making at interim have a different interpretation depending on the three metrics we consider (Bayesian predictive power for a flat prior and two versions of the conditional power). To obtain comparable results across the metrics, the cut-off values are selected to be the 10\%, 15\%,$\ldots$, 50\% percentiles of the different calculated metrics for the longitudinal approaches under all scenarios with an interim analysis at 30\% of the patients. The completer only analyses use the same cut-off as the corresponding longitudinal analysis.

Under the null setting, the measure of interest is the probability to stop the trial for futility. For a pre-specified cut-off value $v$, the futility probability based on data at interim timing $T$ is calculated as:
$$ P(stop\, at\, interim\, T\, |\, v) = \frac{1}{R} \sum_{r=1}^{R} I(p^{(r)}<v), $$
where $I(\cdot)$ is the indicator function, $R$ is the total number of replications implemented in the simulation and $p^{(r)}$ is the predictive or conditional power derived in the $r^{th}$ replication.

Under the alternative setting, it's more of interest to explore the power loss introduced by the interim analysis. The power loss is defined as the probability of stopping the trial for futility at interim when the trial would have succeeded had it continued. In the case with only one interim analysis at time $t$, 
$$Power\,Loss = P(stop\, at\, interim\, t\, and\, final\, analysis\, is\, positive\, |\, v ) = \frac{1}{R} \sum_{r=1}^{R} I(p^{(r)}<v)\times I(\max_{m\in \{1,\cdots,M\}} T_m>c_{1-\alpha}), $$
where $\mathbf{T}=(T_1,\cdots,T_M)$ is the vector of statistics based on all data at the final analysis and $c_{1-\alpha}$ is the corresponding critical value at level $\alpha$.

\subsubsection{Results}

\begin{figure}
    \centering
    \includegraphics[width=0.6\linewidth]{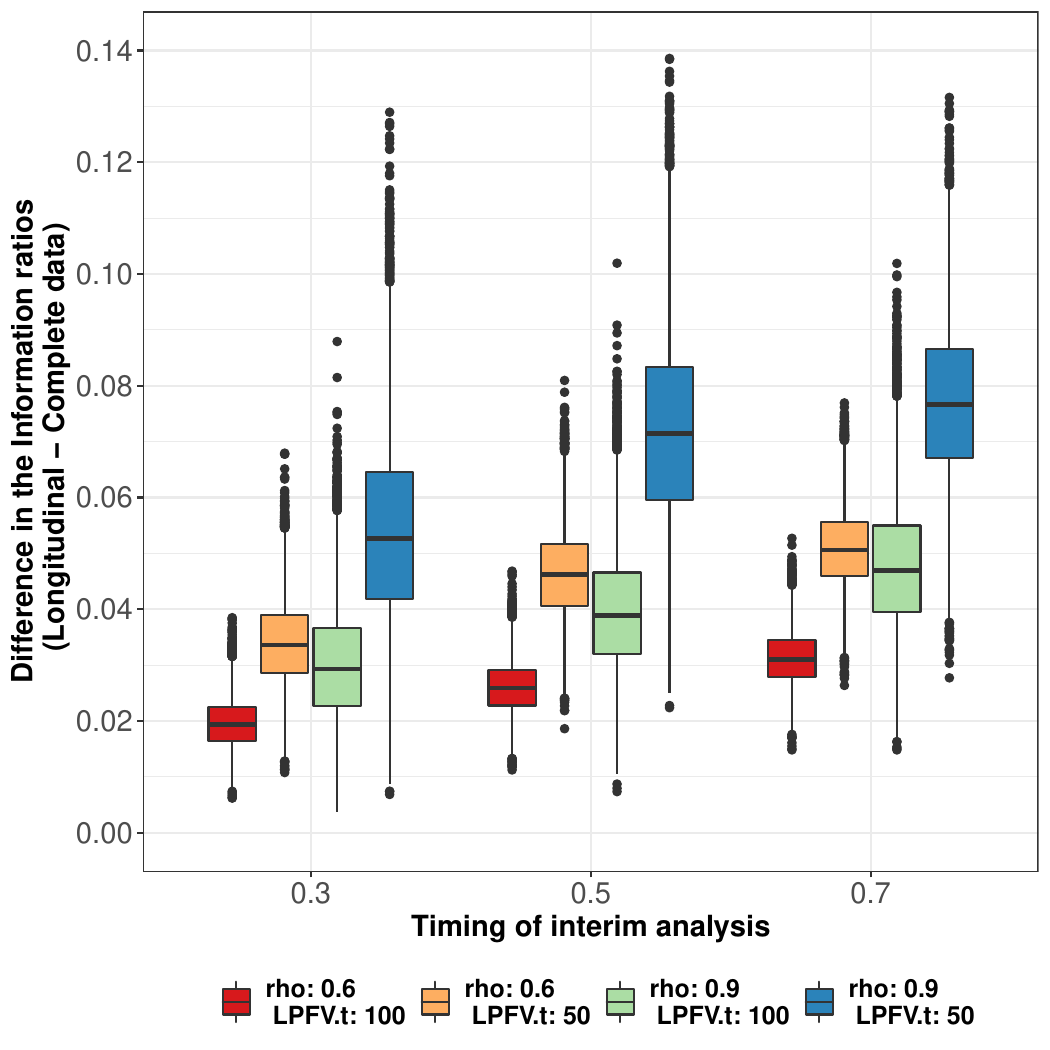}
    \caption{Difference in the Information fractions comparing analysis using all available longitudinal data versus using completers' data only for different scenarios of rho and LPFV.t.}
    \label{fig:info_frac}
\end{figure}

Figure \ref{fig:info_frac} shows the observed difference in the information fractions (\ref{eqn:det}) for the adjusted means at interim using the longitudinal analysis and the completers only. Since the information fraction does not depend on the treatment effect, the difference is summarized based on the combined results under the null and alternative hypothesis for each scenario. As expected the information gain is larger for the longitudinal analysis, when the recruitment speed is higher (i.e. LPFV.t smaller) and the correlation over time is larger. On average the information gain from utilizing the additional data from the incomplete patients ranges between 2\% to 8\% in the scenarios we consider.

\begin{figure}
    \centering
    \includegraphics[width=0.7\linewidth]{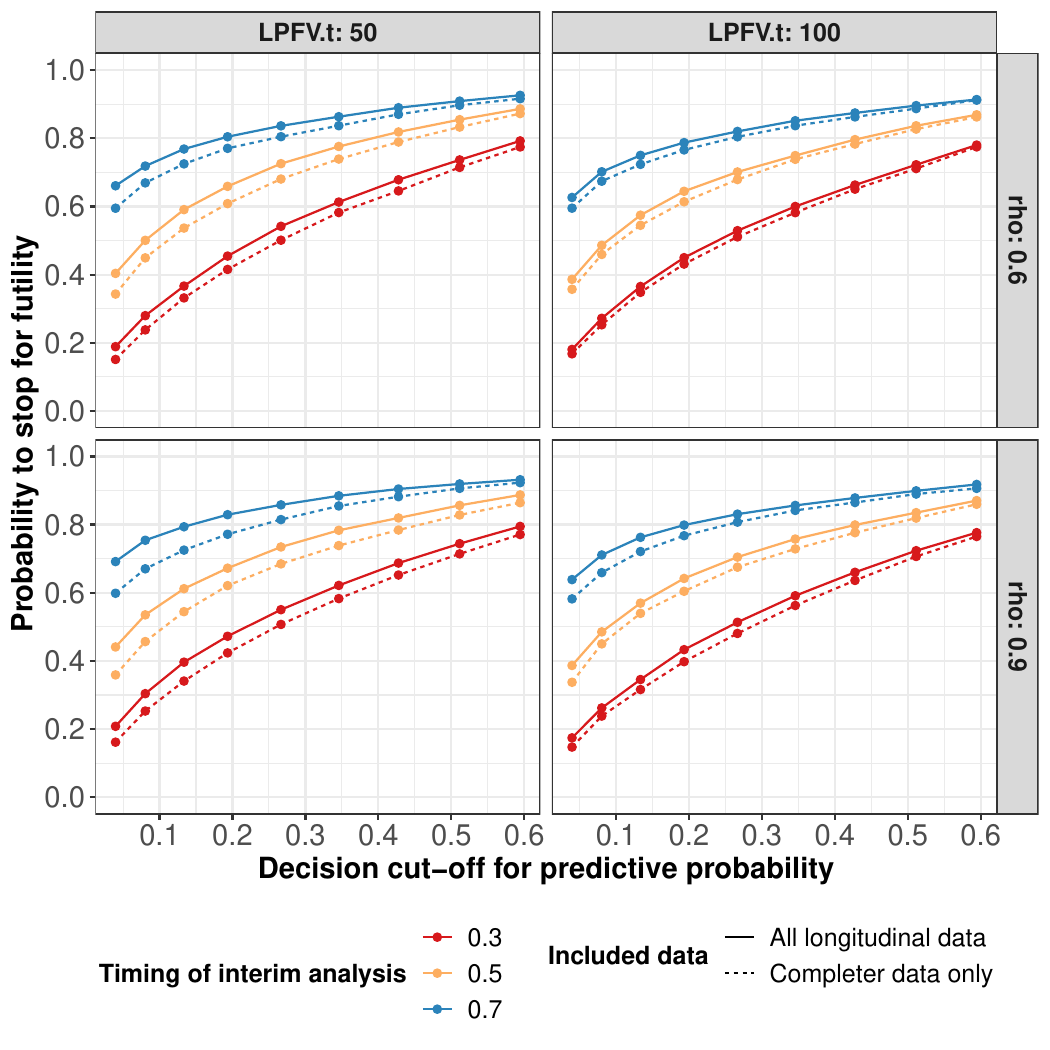}
    \caption{Simulation result: probability to stop for futility (based on predictive power) comparing analysis using all available longitudinal data versus using completers' data only at different timing and for different scenarios of rho and LPFV.t.}
    \label{fig:futility_pp}
\end{figure}

This general trend is also seen in Figure \ref{fig:futility_pp} showing the probability to stop for futility under the null hypothesis if decision is made based on the predictive power. Generally the probability to stop for futility increases when increasing the threshold required on the predictive power to continue. Furthermore the probability to stop for futility increases with more information available towards the end of study. As expected based on the information fractions, the predictive power based on a longitudinal model outperforms the approach based on completers only. The benefit appears larger when the recruitment is faster. Comparing across the different scenarios for $\rho$, the benefit of a longitudinal approach seems to be only slightly larger when $\rho=0.9$.

\begin{figure}[h]
    \centering
    \includegraphics[width=0.7\linewidth]{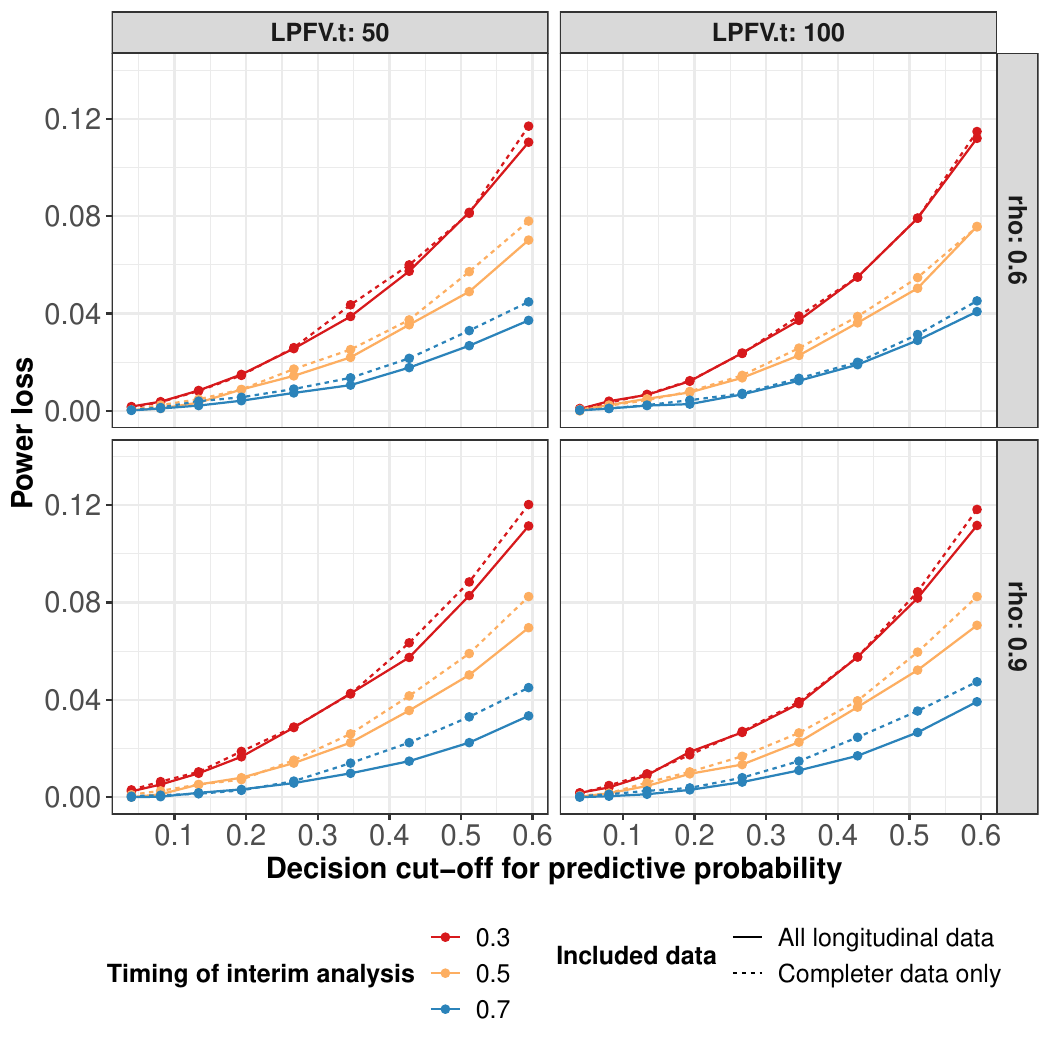}
    \caption{Simulation result: power loss (based on predictive power) comparing analysis using all available longitudinal data versus using completers' data only at different timing and for different scenarios of rho and LPFV.t.}
    \label{fig:powloss_pp}
\end{figure}

In terms of the power loss in Figure \ref{fig:powloss_pp} under the scenario when the maximum effect is 0.12 at the 8mg dose at week 12, qualitatively similar results are obtained. The power loss gets larger with increasing threshold, in addition the power loss gets larger when the interim analysis is conducted earlier. The power loss based on the longitudinal analysis is smaller or equal to the power loss based on an analysis based on completers alone. The benefit of the longitudinal analysis seems to be larger when the recruitment is faster. Again it seems that when $\rho$ is larger the benefit of the longitudinal analysis is also slightly larger.

In conclusion, by incorporating additional incompleters' data in longitudinal analysis, benefits were observed in terms of increased chances of early stopping a futile study and avoiding early termination of a positive study. The relative gain appears to be consistent with the gain observed in the information fraction, which is affected by the recruitment speed and outcome correlations.

Qualitatively very similar simulation results hold for the two versions of conditional power and are reported in the Appendix.

\section{Example application}
\label{sec:example}

In this section we will revisit the study introduced in Section \ref{sec:intro_example}. The study was stopped early for reasons unrelated to data observed in the trial, when around half of the originally planned sample size was recruited for the study. The study did not prospectively plan for a futility analysis. In this section we would like to use the data from the study to retrospectively perform futility analyses at different time points throughout the trial, to illustrate the methodologies presented in the previous sections.

Analyses will be performed at 7 time points, where at each time-point only data collected before the analysis time point are included. The time-points will be selected based on the percentage of patients who have completed the week 12 measurement. We will consider 7 equally spaced values ranging from 40\% to 100\% of the recruited patients in the data (corresponding to roughly 20\% to 50\% of the originally planned sample size). The primary endpoint was average change from baseline in pre-dose FEV1 (L) at Week 8 and Week 12. For simplicity, we used change from baseline in pre-dose FEV1 (L) at Week 12 as the endpoint here for the purpose of this paper. The longitudinal model was fitted with a linear mixed effects model for repeated measures (MMRM) with fixed effects including baseline FEV1, visit (week 2, 4, 8 and 12), treatment, randomization strata (EOS $\geq$ 300 or < 300 cells/µl), region, treatment-by-visit interaction, baseline FEV1-by-visit interaction, FEV1 prior to inhalation, and FEV1 within 30 min post inhalation of salbutamol/albuterol (components of SABA reversibility). To allow adjustment for correlations between time points within patients, an unstructured variance-covariance structure was used. An analysis of covariance (ANCOVA) model was fitted for the completer model, with baseline FEV1, treatment, randomization strata (EOS $\geq$ 300 or < 300 cells/µl), region, FEV1 prior to inhalation, and FEV1 within 30 min post inhalation of salbutamol/albuterol (components of SABA reversibility) as covariates. 

For each time-point we will evaluate the information fraction from completer and longitudinal analysis using the formula in Equation (\ref{eqn:det}). In addition we will calculate the Bayesian predictive power (corresponding to a non-informative prior) and the conditional power, where the originally planned treatment effect is assumed for the second stage of the trial as described in Section \ref{sec:methodology}.

\begin{figure}[h]
    \centering
    \includegraphics[width=0.7\linewidth]{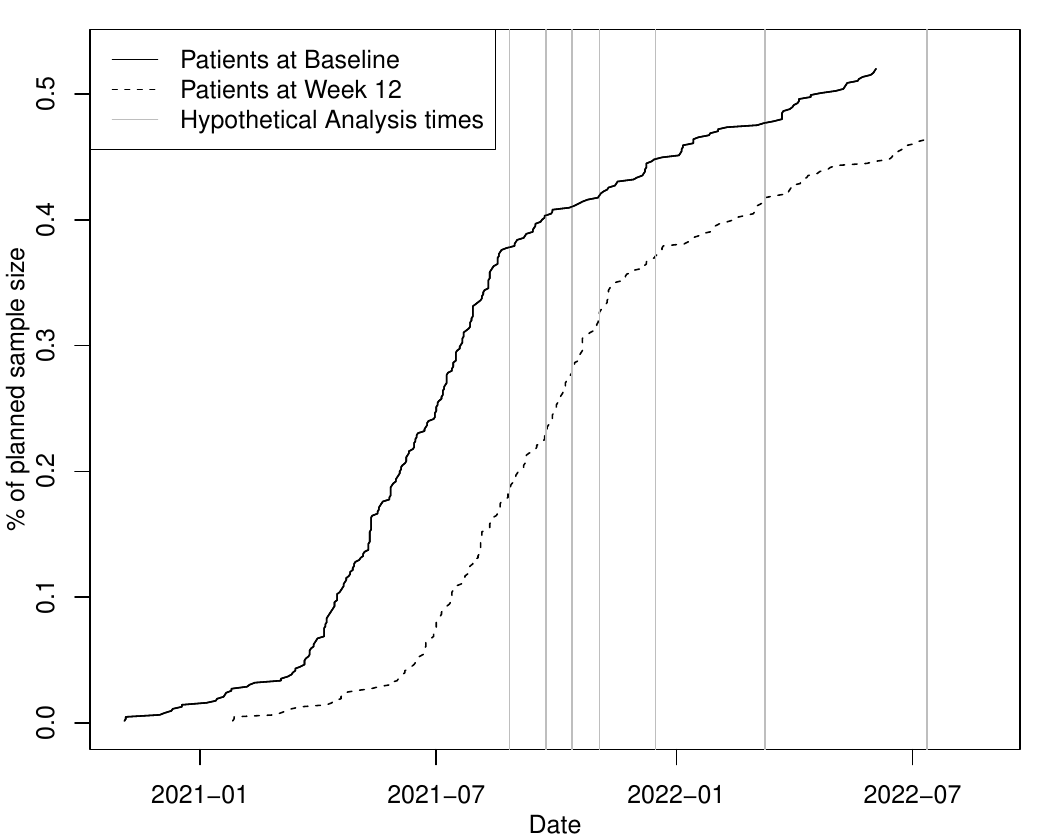}
    \caption{Recruitment pattern from the example study}
    \label{fig:example_recruit}
\end{figure}

Figure \ref{fig:example_recruit} shows the recruitment pattern as observed in the trial. After an initial ramp-up phase, recruitment accelerated, but then after a certain time-point also slowed down again slightly. Note that the dashed line (indicating patients completing the week 12 visit) is lower than the number of patients recruited at the last time-point indicating that some patients were lost to follow-up before week 12. The vertical lines in the plot indicate the the 7 time-points, where the interim analyses were performed for the purpose of this paper.

\begin{figure}[h]
    \centering
    \includegraphics[width=0.7\linewidth]{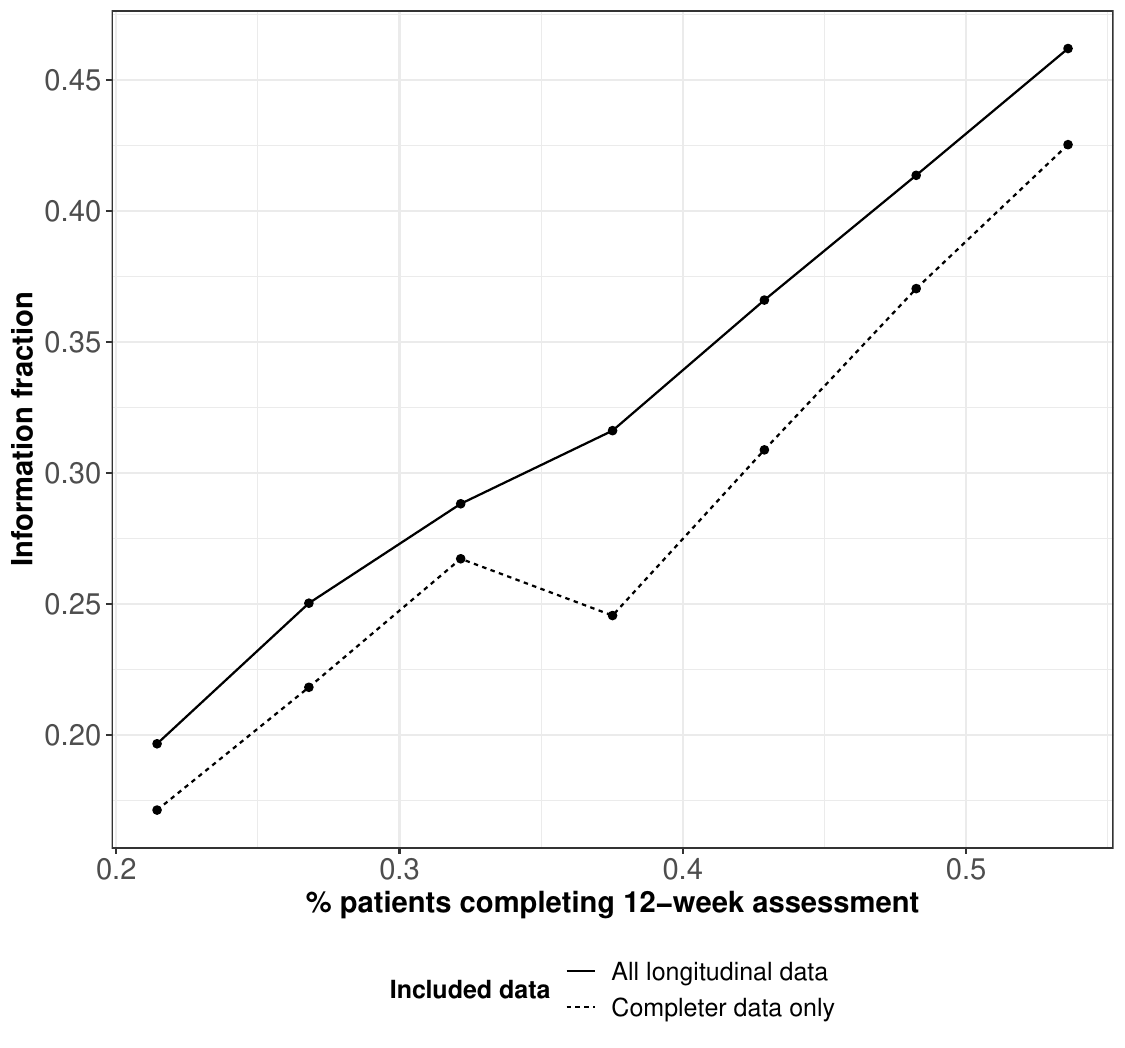}
    \caption{Information fraction at different interim analysis time-points from the example study, using all available longitudinal data versus using completers' data only}
    \label{fig:example_inf_frac}
\end{figure}

Figure \ref{fig:example_inf_frac} displays the information fraction. As can be seen the longitudinal analysis over time leads to an increase in the information fraction of around 3-4\%. This is similar to the simulation scenario of rho = 0.9 and LPFV.t = 100 in Figure \ref{fig:info_frac}, which most closely reflects the underlying data. Of note, at the fourth interim analysis, there was a dip in the information fraction for the completer model, which can be explained by the fact that both numerator and denominator in the information fraction formula are updated at each time point. 

\begin{figure}[h]
    \centering
    \includegraphics[width=0.7\linewidth]{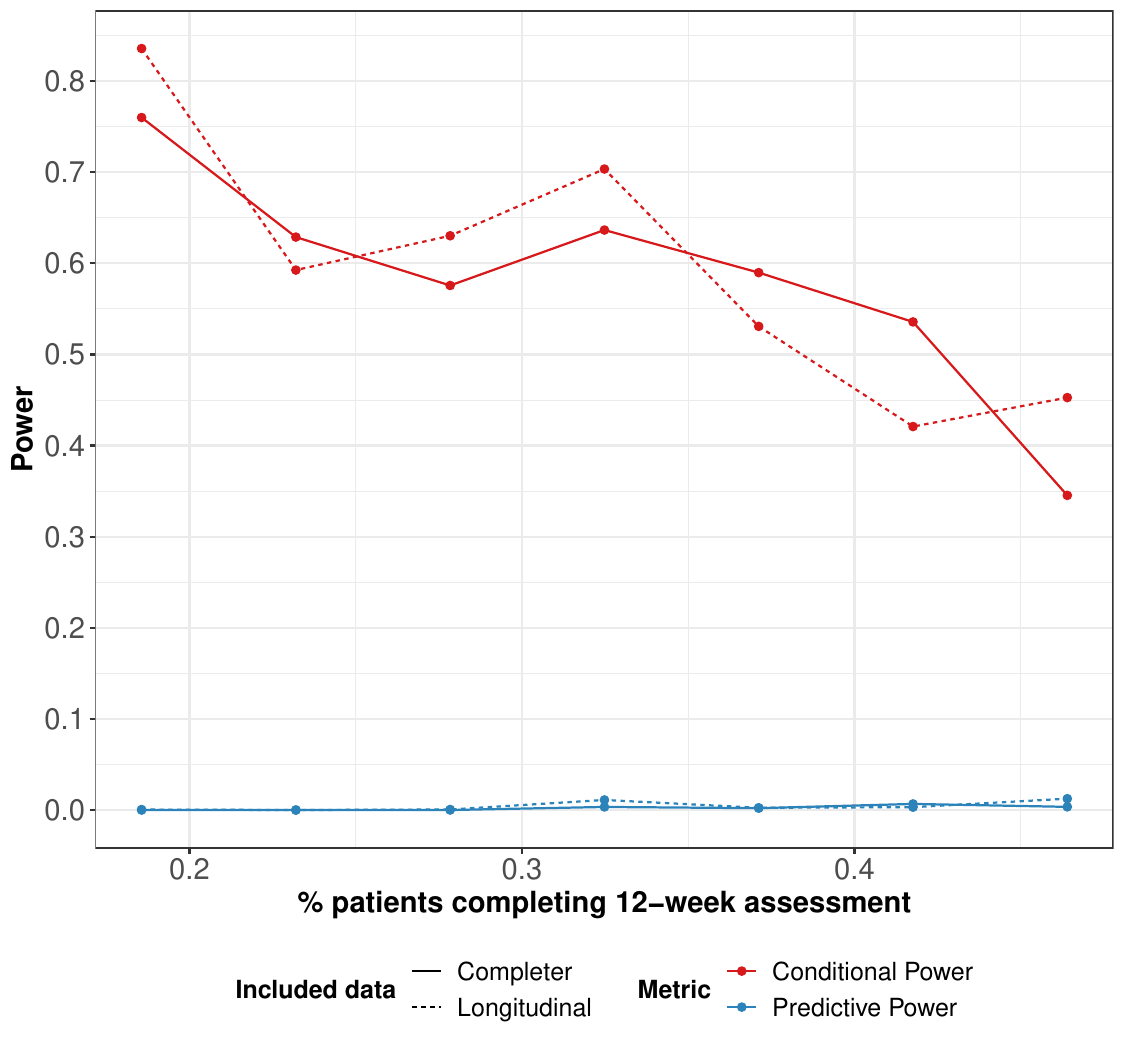}
    \caption{Predictive and conditional powers at different interim analysis time-points from the example study, using all available longitudinal data versus using completers' data only}
    \label{fig:example_PP_CP}
\end{figure}

Figure \ref{fig:example_PP_CP} shows the predictive and conditional power over time using all available longitudinal data versus using completers' data only. The conditional power gets lower with increasing interim analysis time, which is due to the fact that the design based assumption has less and less impact on the conditional power. The predictive power is close to 0 even for the early analyses. This is because it is only based on the data from the trial (not based on design assumptions). Note that both metrics reflect quite different information: The conditional power for the planned effect assumes that for part 2 of the trial the originally planned assumptions are correct, while the predictive power uses the information learned from the first part of the data. For the simulation study the difference in scaling is also visible in the x-axis limits in Figures \ref{fig:futility_pp}, \ref{fig:powloss_pp}, \ref{fig:futility_cp}, \ref{fig:powloss_cp}. In the simulation settings controlling the power-loss at around 2\% values required cut-off values between 20\% and 50\% (depending on time-point of interim analysis and scenario) for the predictive power, while for the conditional power based on the planned effect, larger cut-offs are required (between 50\% and 60\%). Based on these cut-offs, both metrics indicate a futility conclusion early during the trial, the predictive power very early while the conditional power a bit later.

\section{Discussion and Conclusions}
\label{sec:concl}

Futility analyses provide a risk mitigation strategy in Phase 2 trials which are gatekeepers to often costly and time-consuming Phase 3 trials. While MCP-Mod is commonly used to design and analyze randomized Phase 2 trials, its performance is less investigated in combination with commonly used futility criteria such as predictive and conditional power. This paper proposes a general framework to allow futility decisions based on data available at the interim analysis with a focus on longitudinal models to utilize incomplete data. Through extensive simulations, it is demonstrated that the completers analysis (without using incomplete longitudinal data) often has less favorable operating characteristics. The gain of utilizing longitudinal data is more obvious when the recruitment speed is high and when the correlation over time is larger.

One critical aspect of determining futility criteria such as predictive and conditional power is to calculate the information fraction at the interim analysis. This paper proposes to calculate information using the determinant of the covariance matrix of adjusted dose group means and to decompose the total information as the sum of information by and after the interim analysis. The former can be estimated from the longitudinal data available at the interim analysis including patients with missing outcomes for the primary time-point of interest. The latter requires projection of how many patients with outcomes at the primary time-point of interest and potentially with outcomes at other times, if some patients fail to complete the study due to, for example, loss to follow-up. This missing data issue raise challenges of reliably estimating the information at the final analysis. One possible solution is to follow the missing data pattern observed at the interim analysis. For example, among those who potentially should complete the study at the interim analysis, we could obtain the pattern of data availability at the primary time-point of interest and other times, and assume the same pattern at the final analysis.

From a methodological perspective, futility analysis thresholds can be defined on different scales, for example also based on point estimates or $p$-values at interim. In this paper we focused on predictive or conditional power metrics. These metrics could be more adequate in terms of communication as they make statements about the likelihood of success of the completed trial, which is more closely aligned to the purpose of futility analyses.  When comparing predictive and conditional power it has been established in the setting with two treatment groups that one metric can be converted to the other metric for a single interim analysis \cite{gallo2014alternative}. This means that the futility decision using one metric and a cut-off value can be equivalently expressed using the other metric with a different cut-off value. This relationship should remain the case for multiple dose groups using the generalized MCP-Mod approach because of the monotonic relationship between the cut-off value of the probability to stop the trial for futility. This again shifts the focus of futility decisions from the metric and the cut-off value to their operating characteristics, i.e., the stopping probability under the null hypothesis and the power loss under the alternative hypothesis.

Note that we primarily focused on futility analyses in this paper, while the methods could equally be used for unblinded sample size re-estimation.

%\backmatter

\textbf{Acknowledgments}\\
We would like to thank Steffen Ballerstedt for his support and feedback on an earlier version of this manuscript

%\bmsection*{Financial disclosure}

%None reported.

%\bmsection*{Conflict of interest}

%The authors declare no potential conflict of interests.

\bibliographystyle{abbrv}

\appendix

\section{Illustration of benefit for longitudinal analysis}
\label{sec:illus}

Consider the setting of a single arm study where three post-baseline measurements, equally spaced over time, are taken for each patient. Of interest is the group mean at the last visit. We assume that the patient specific data are generated from a multivariate normal distribution with a given mean vector over time and a compound symmetry covariance structure. We assume the standard deviation of the residual error is 1 at each visit and the equal correlation across the visits is $\rho$. 

Suppose for illustrative purposes we recruit 21 patients in the trial. At the end of the study the information (inverse variance) of the group mean estimator is hence given by $21$.  At an interim analysis the information for the group mean estimate at interim for an analysis that only uses data from the final visit (completers) is given by the number of patients that have completed the third post-baseline visit. For the longitudinal model that uses all post-baseline visits one can fit a multivariate normal model. The information (inverse variance), can be explicitly calculated based on generalized least squares, by calculating the inverse of the covariance matrix $(\mathbf{X}'\boldsymbol{\Sigma}^{-1}\mathbf{X})^{-1}$. Here $\mathbf{X}$ is a matrix which has $3$ columns and as many rows as there are measurements at the time of interim analysis. The matrix $\boldsymbol{\Sigma}$ is a block-diagonal with as many rows as there are observations. The diagonal elements are patient specific $3\times 3$ matrices with only ones on the diagonal and $\rho$ on the off-diagonal (for patients who have not completed the final visit the matrix will be of smaller dimension). The variance of the group mean for the third visit can be extracted from this covariance matrix.

One main factor that impacts the comparison of using all data based on a longitudinal model versus using the completers is how many more incomplete patients are available at interim. This in turn is impacted by the recruitment speed and the recruitment pattern. A further factor is the correlation of the earlier measurements with the measurement at visit 3. 

\begin{figure}[h]
    \centering
    \includegraphics[width=0.5\linewidth]{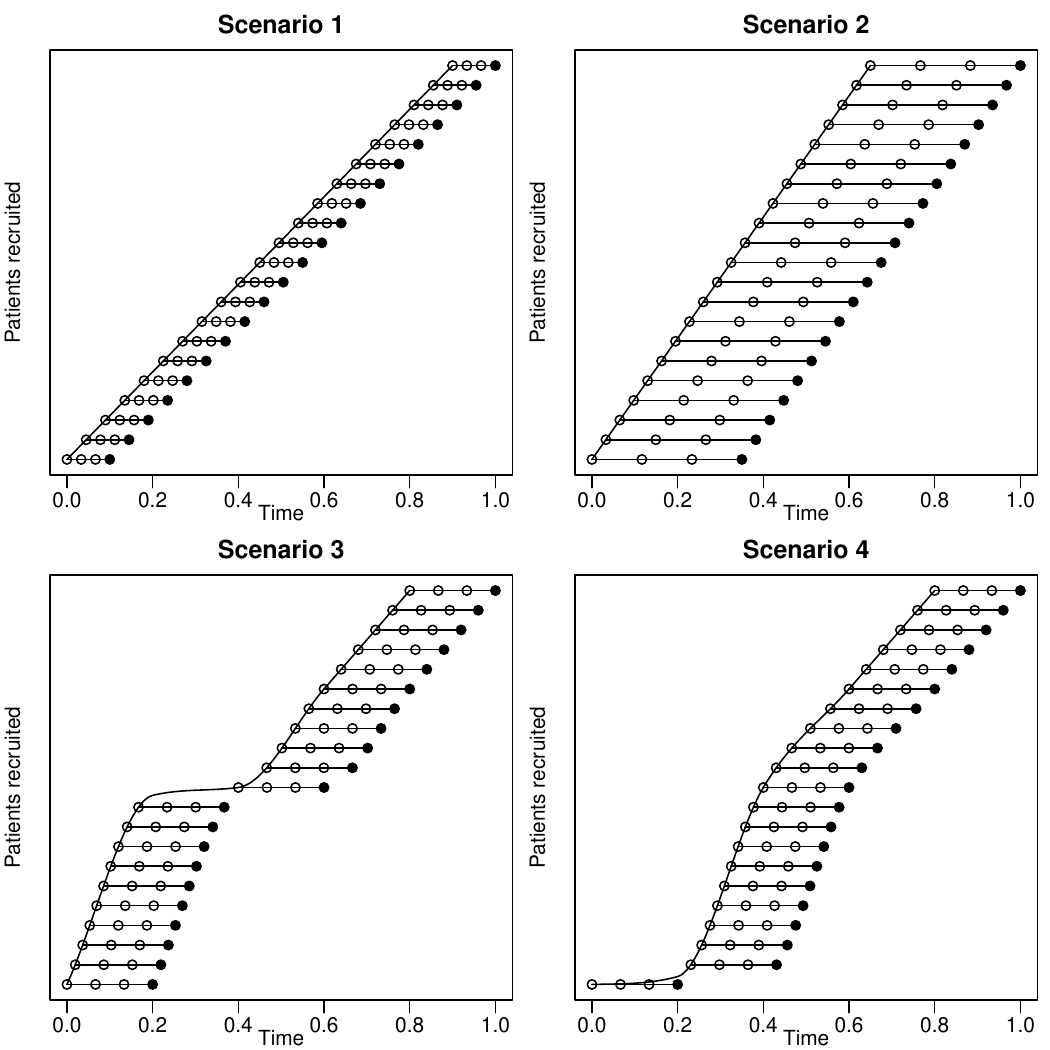}
    \caption{4 Scenarios on how the 21 patients are recruited into the trial. Each visit (1 baseline and 3 post-baseline) are indicated by a circle for each patient. The filled circle corresponds to the final visit.}
    \label{fig:recruitment_scen}
\end{figure}

To illustrate these factors we consider the four scenarios in Figure \ref{fig:recruitment_scen}. The difference between Scenario 1 and 2 is the speed of recruitment, that is how fast is the recruitment of all patients relative to the time it takes to assess the endpoint for one patient. In Scenario 1 the recruitment is slower: Recruitment takes much longer than the time to measure the final endpoint for one patient. For Scenario 2 the recruitment is faster. For Scenarios 3 and 4 the overall speed of recruitment is the same, but the recruitment pattern is different.

\begin{figure}[h]
    \centering
    \includegraphics[width=0.5\linewidth]{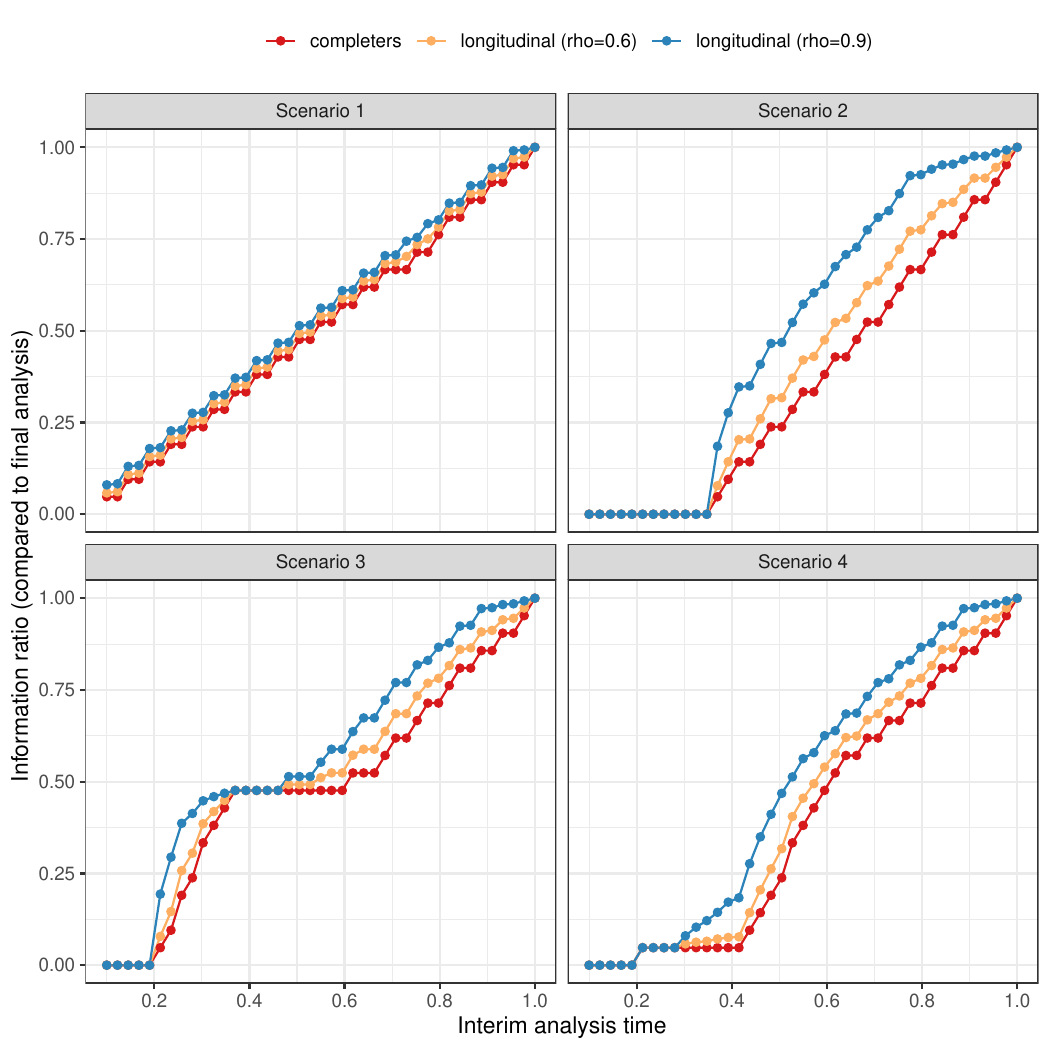}
    \caption{Information fraction at different possible interim analysis time-points, compared to the information at study end. The completer analysis is compared against a longitudinal analysis using all post-baseline measurement.}
    \label{fig:info_scenarions}
\end{figure}

In Figure \ref{fig:info_scenarions} one can observe how the information fraction for the longitudinal analysis (assuming the correlation is $0.6$ or $0.9$) and the completer analysis for different time-points of performing the interim analysis. As expected the longitudinal analysis always provides larger information (a smaller standard error) compared to the completer analysis. In particular when the correlation is higher. The benefit is larger in Scenario 2 compared to Scenario 1. The reason for that is that at interim analysis there are much more patients with incomplete data. These can be included in the longitudinal analysis, but not in the completer analysis. The results for Scenario 3 and 4 also illustrate this general trend: For Scenario 3 an interim analysis at time 0.4 the completer and longitudinal analysis will contain the same information, due to the unusual recruitment shape (see Figure \ref{fig:recruitment_scen}). However for an interim analysis at time point 0.5 for Scenario 4, there will be many patients with incomplete data, so the longitudinal analysis will contain relatively more information compared to a completer analysis.

We believe similar trends also hold when one considers treatment comparisons: The variance of the treatment contrast estimate will then be the weighted sum of the variance of the specific dose groups, so a similar patterns are expected to appear for situations with more than one treatment group, such as general contrasts used for MCP-Mod.
\newpage

\section{Additional simulation results}
\label{sec:addres}

\begin{figure}[h]
    \centering
    \includegraphics[width=0.6\linewidth]{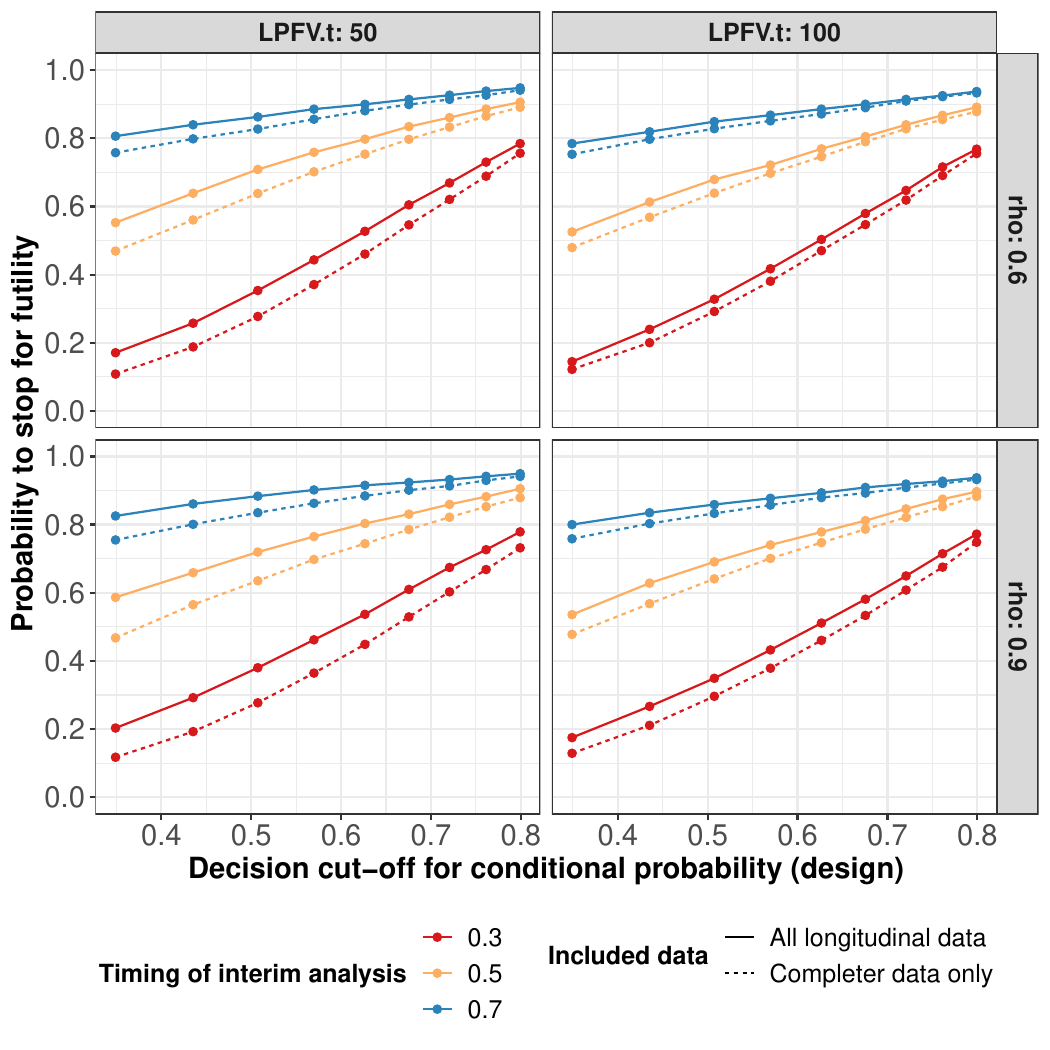}
    \caption{Simulation result: probability to stop for futility (based on conditional power calculated using assumed treatment effects) comparing analysis using all available longitudinal data versus using completers' data only at different timing and for different scenarios of rho and LPFV.t.}
    \label{fig:futility_cp}
\end{figure}

\begin{figure}[h]
    \centering
    \includegraphics[width=0.6\linewidth]{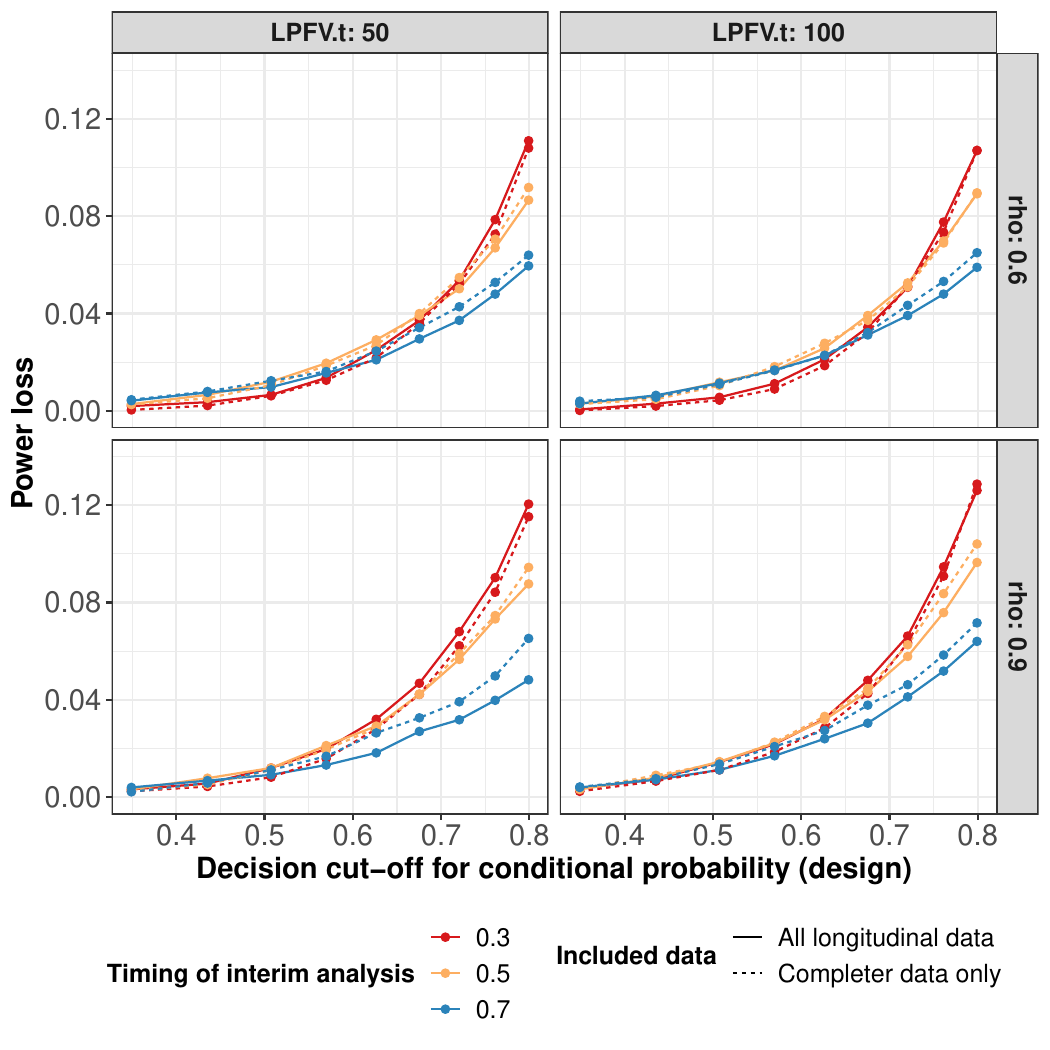}
    \caption{Simulation result: power loss (based on conditional power calculated using assumed treatment effects) comparing analysis using all available longitudinal data versus using completers' data only at different timing and for different scenarios of rho and LPFV.t.}
    \label{fig:powloss_cp}
\end{figure}

\begin{figure}[h]
    \centering
    \includegraphics[width=0.6\linewidth]{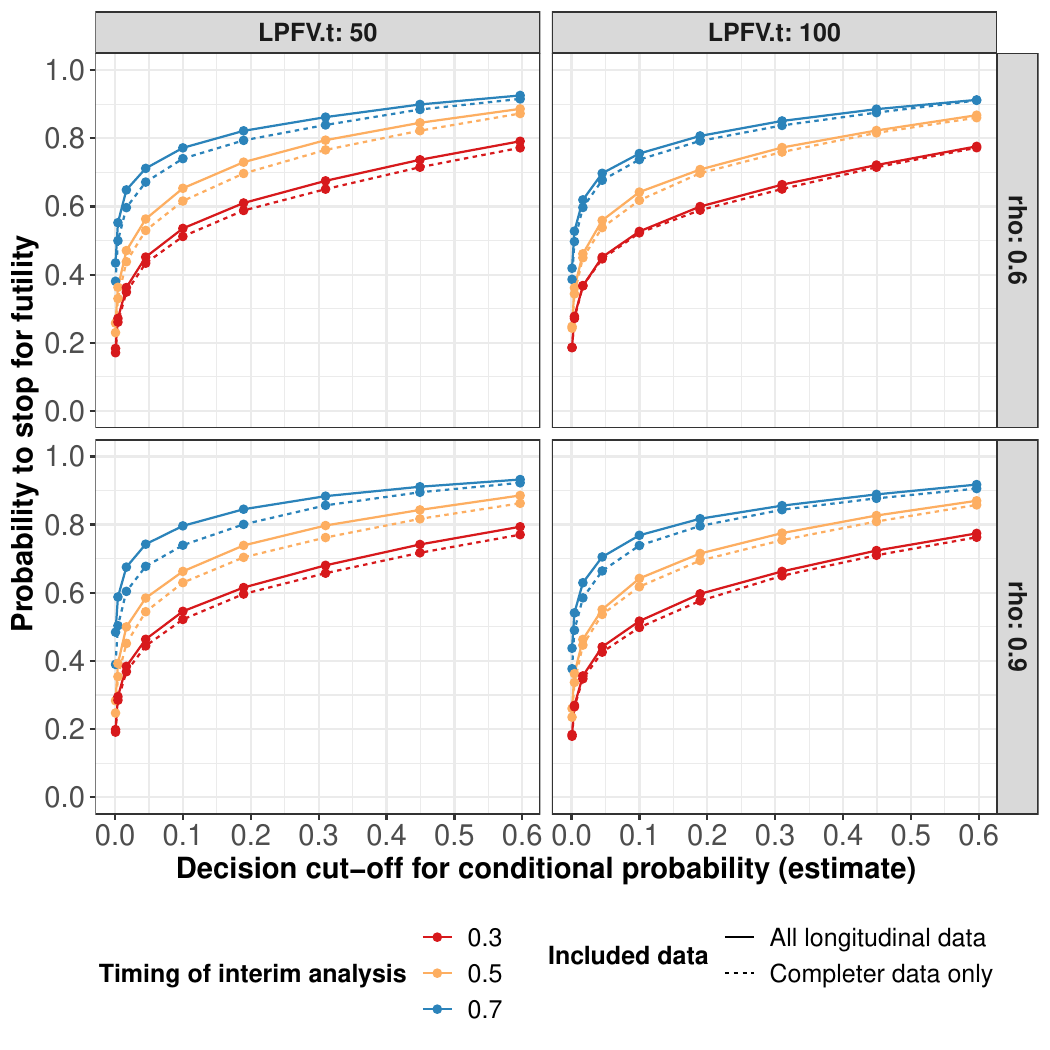}
    \caption{Simulation result: probability to stop for futility (based on conditional power calculated using estimated means at interim) comparing analysis using all available longitudinal data versus using completers' data only at different timing and for different scenarios of rho and LPFV.t.}
    \label{fig:futility_cp2}
\end{figure}

\begin{figure}[h]
    \centering
    \includegraphics[width=0.6\linewidth]{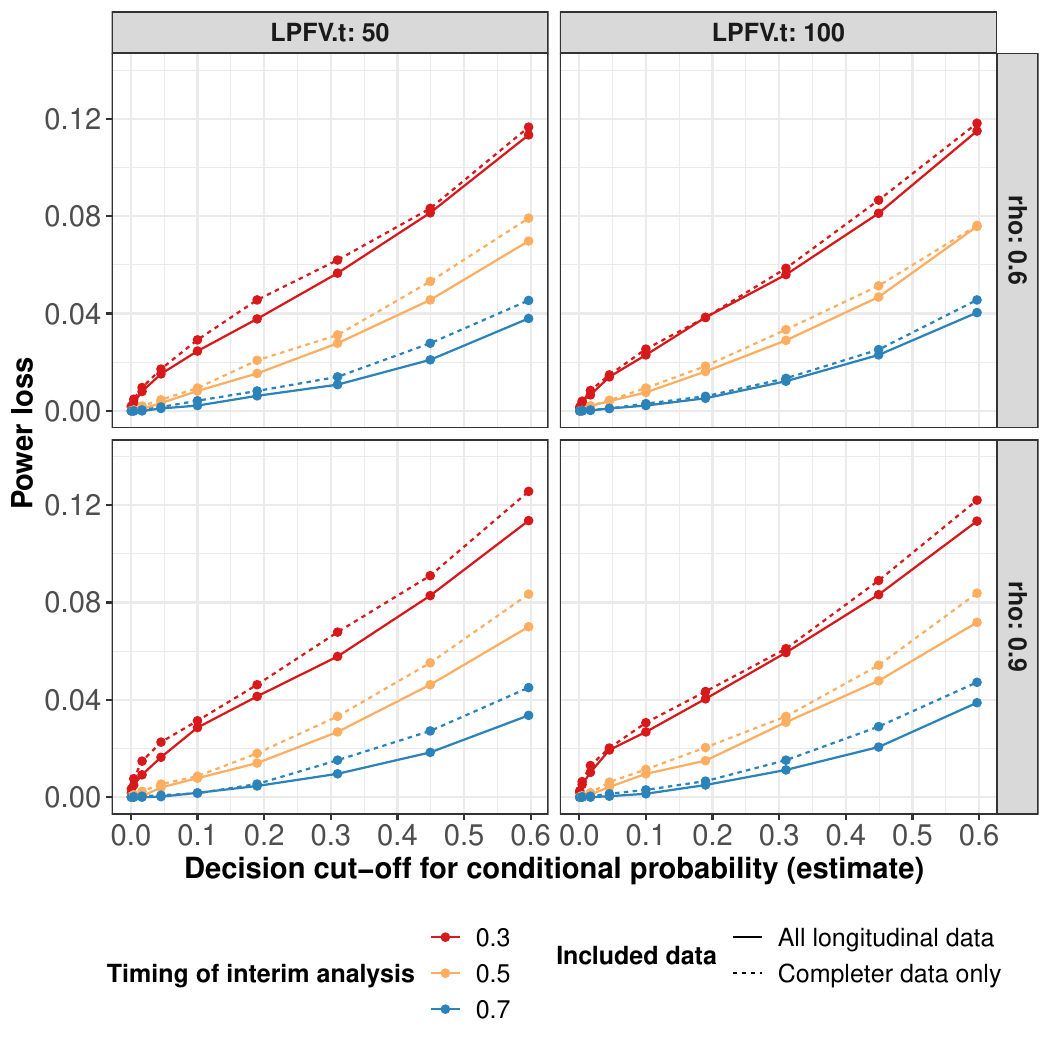}
    \caption{Simulation result: power loss (based on conditional power calculated using estimated means at interim) comparing analysis using all available longitudinal data versus using completers' data only at different timing and for different scenarios of rho and LPFV.t.}
    \label{fig:powloss_cp2}
\end{figure}

\end{document}